\documentclass[final,3p,times,twocolumn]{elsarticle}

 \usepackage{graphicx}
\usepackage{amssymb}
\usepackage{amsmath}
\usepackage{hyperref} 
\usepackage{longtable}
\biboptions{sort&compress}

\journal{Physics Letters B}

\newcommand{\preprint}{
 \setlength{\unitlength}{1mm}{\hbox{\begin{picture}(0,0)
      \put(160,10){\mbox{\footnotesize%
        ADP-09-10/T688}}\end{picture}}}}

\begin{document}

\begin{frontmatter}

\title{\preprint 
       Isolating the Roper Resonance in Lattice QCD}

\author[adl,raj]{M.S. Mahbub}
\author[adl,cyp]{Alan $\acute{\rm{O}}$ Cais}
\author[adl]{Waseem Kamleh}
\author[adl]{Ben G. Lasscock}
\author[adl]{Derek B. Leinweber}
\author[adl]{Anthony G. Williams}
\address[adl]{Special Research Centre for the Subatomic Structure of Matter, Adelaide, South Australia 5005, Australia, \\ and Department of Physics, University of Adelaide, South Australia 5005, Australia.}
\address[raj]{Department of Physics, Rajshahi University, Rajshahi 6205, Bangladesh.}
\address[cyp]{Cyprus Institute, Guy Ourisson Builiding, Athalassa Campus, PO Box 27456, 1645 Nicosia, Cyprus.}

\begin{abstract}

We present results for the first positive parity excited state of the
nucleon, namely, the Roper resonance ($N^{{\frac{1}{2}}^{+}}$=1440
MeV) from a variational analysis technique. The analysis is performed
for pion masses as low as 224 MeV in quenched QCD with the FLIC fermion
action. A wide variety of smeared-smeared correlation functions are
used to construct correlation matrices. This is done in order to find
a suitable basis of operators for the variational analysis such that
eigenstates of the QCD Hamiltonian may be isolated. A lower lying Roper
state is observed that approaches the physical Roper state.
 To the best of our knowledge, the first time this state
has been identified at light quark masses using a variational approach.
\end{abstract}


\end{frontmatter}

One of the long-standing puzzles in hadron spectroscopy has been the
low mass of the first positive parity, $J^{P}={\frac{1}{2}}^{+}$,
excitation of the nucleon, known as the Roper resonance $N^{*}$(1440
MeV). In constituent or valence quark models with harmonic oscillator
potentials, the lowest-lying odd parity state naturally occurs below
the $N={\frac{1}{2}}^{+}$ state (with principal quantum number $N=2$)
~\cite{Isgur:1977ef,Isgur:1978wd} whereas, in nature the Roper
resonance is almost 100 MeV below the $N={\frac{1}{2}}^{-}$(1535 MeV)
state. Similar difficulties in the level orderings appear for the
$J^{P}={\frac{3}{2}}^{+} \Delta^{\ast}(1600)$ and ${\frac{1}{2}}^{+}
\Sigma^{\ast} (1690)$ resonances, which have led to the speculation
that the Roper resonance may be more appropriately viewed as a hybrid
baryon state with explicitly excited gluon field configurations
~\cite{Li:1991yba,Carlson:1991tg}, or as a breathing mode of the
ground state ~\cite{Guichon:1985ny} or states which can be described
in terms of meson-baryon dynamics alone ~\cite{Krehl:1999km}.

The
first detailed analysis of the positive parity excitation of the nucleon
was performed in Ref.~\cite{Leinweber:1994nm} using Wilson fermions
and an operator product expansion spectral ansatz. Since then several
attempts have been made to address these issues in the lattice
framework
~\cite{Lee:1998cx,Gockeler:2001db,Sasaki:2001nf,Melnitchouk:2002eg,Edwards:2003cd,Lee:2002gn,Mathur:2003zf,Sasaki:2003xc,Mahbub:2009nr,Guadagnoli:2004wm},
but in many cases no potential identification of the Roper state has
been
made~\cite{Lee:1998cx,Gockeler:2001db,Sasaki:2001nf,Melnitchouk:2002eg,Edwards:2003cd}. Recently
however, in the analysis of ~\cite{Lee:2002gn,Sasaki:2005ap,Mathur:2003zf}, a
low-lying Roper state has been identified using Bayesian
techniques.\\ 

Here, we use a `variational
method' ~\cite{Michael:1985ne,Luscher:1990ck,McNeile:2000xx}, which is
based on a correlation matrix analysis and has been used quite
extensively in
Refs.~\cite{Blossier:2009kd,Allton:1993wc,McNeile:2000xx,Melnitchouk:2002eg,Hedditch:2003zx,Lasscock:2007ce,Brommel:2003jm,Burch:2004he,Burch:2004zx,Burch:2005vn,Lasscock:2005tt,Burch:2005wd,Burch:2006dg,Burch:2006cc,Burch:2005md,Burch:2005qf,Basak:2006ki,Basak:2007kj,Mahbub:2009nr}. Though
the ground state mass of the nucleon has been described successfully,
an unambiguous determination of the Roper state with this method has not been achieved
in the past, though significant amounts of
research have been carried out in Ref.~\cite{Allton:1993wc}, by the CSSM
Lattice Collaboration
~\cite{Melnitchouk:2002eg,Lasscock:2007ce,Mahbub:2009nr}, the BGR
~\cite{Brommel:2003jm,Burch:2004he,Burch:2004zx,Burch:2005vn,Burch:2006cc}
Collaboration and in Refs.~\cite{Basak:2006ki,Basak:2007kj}.\\
In this Letter, we present evidence of a low-lying Roper state for the
first time using a
variational analysis. The observed state displays chiral curvature and
approaches the physical mass of the Roper state. The standard nucleon
interpolating field $\chi_{1}$ is considered in this analysis. Various
sweeps of Gaussian smearing ~\cite{Gusken:1989qx} are used to construct
a smeared-smeared correlation function basis from which we obtain the
correlation matrices.\\
The two point correlation function matrix for $\vec{p} =0$ can be written as
\begin{align}
G_{ij}(t) &= (\sum_{\vec x}{\rm Tr}_{\rm sp}\{ \Gamma_{\pm}\langle\Omega\vert\chi_{i}(x)\bar\chi_{j}(0)\vert\Omega\rangle\}) \\
          &=\sum_{\alpha}\lambda_{i}^{\alpha}\bar\lambda_{j}^{\alpha}e^{-m_{\alpha}t},
\end{align}
where, Dirac indices are implicit. Here, $\lambda_{i}$ and
$\bar\lambda_{j}$ are the couplings of interpolators $\chi_{i}$ and
$\bar\chi_{j}$ at the sink and source, respectively. $\alpha$
enumerates the energy eigenstates with mass $m_{\alpha}$.\\ 
 Since the only $t$ dependence comes from the exponential term, one
 can seek a linear superposition of interpolators,
 ${\bar\chi}_{j}u_{j}^{\alpha}$, such that (more detail can be found in Refs.~\cite{Melnitchouk:2002eg,Mahbub:2009nr}), 
\begin{align}
G_{ij}(t+\triangle t)\, u_{j}^{\alpha} & = e^{-m_{\alpha}\triangle
  t}\, G_{ij}(t)\, u_{j}^{\alpha},
\end{align}  
for sufficiently large $t$ and $t+\triangle t$, see Refs.
~\cite{Blossier:2009kd} and ~\cite{Mahbub:2009nr}. Multiplying the above equation by $[G_{ij}(t)]^{-1}$ from the left leads to an eigenvalue equation,
\begin{align}
[(G(t))^{-1}G(t+\triangle t)]_{ij}\, u^{\alpha}_{j} & = c^{\alpha}\, u^{\alpha}_{i},
 \label{eq:right_evalue_eq}
\end{align} 
where $c^{\alpha}=e^{-m_{\alpha}\triangle t}$ is the eigenvalue. Similar to Eq.(\ref{eq:right_evalue_eq}), one can also solve the left eigenvalue equation to recover the $v^{\alpha}$ eigenvector,
\begin{align}
v^{\alpha}_{i}\, [G(t+\triangle t)(G(t))^{-1}]_{ij} & = c^{\alpha}v^{\alpha}_{j}.
\label{eq:left_evalue_eq}
\end{align} 
The vectors $u_{j}^{\alpha}$ and $v_{i}^{\alpha}$  diagonalize the correlation matrix at time $t$ and $t+\triangle t$ making the projected correlation matrix,
\begin{align}
v_{i}^{\alpha}G_{ij}(t)u_{j}^{\beta} & \propto \delta^{\alpha\beta}.
 \label{projected_cf} 
\end{align} 
The parity projected, eigenstate projected correlator, 
\begin{align}
 v_{i}^{\alpha}G^{\pm}_{ij}(t)u_{j}^{\alpha} & \equiv G^{\alpha}_{\pm},
 \label{projected_cf_final}
\end{align}
 is then analyzed using standard techniques to obtain masses of different states.

Our analysis is exploratory, seeking to develop techniques to access
the Roper state in lattice gauge theory. Our lattice ensemble consists of 200 quenched configurations with a lattice
volume of $16^{3}\times 32$. Gauge field configurations are generated
by using the DBW2 gauge action
~\cite{Takaishi:1996xj,deForcrand:1999bi} and an
${\cal{O}}(a)$-improved FLIC fermion action ~\cite{Zanotti:2001yb} is
used to generate quark propagators. This action has excellent scaling
properties and provides near continuum results at finite lattice
spacing ~\cite{Zanotti:2004dr}. The lattice spacing is $a=0.1273$ fm,
as determined by the static quark potential, with the scale set using
the Sommer scale, $r_{0}=0.49$ fm ~\cite{Sommer:1993ce}. In the
irrelevant operators of the fermion action we apply four sweeps of
stout-link smearing to the gauge links to reduce the coupling with the
high frequency modes of the theory ~\cite{Morningstar:2003gk}
providing ${\cal O}(a)$ improvement ~\cite{Zanotti:2004dr}. We use
the same method as in Ref.~\cite{Lasscock:2005kx,Mahbub:2009nr} to determine fixed
boundary effects, and the effects are significant only after time
slice 25 in the present analysis. Various
sweeps (1, 3, 7, 12, 16, 26, 35, 48 sweeps corresponding to rms
  radii, in lattice units, of 0.6897, 1.0459, 1.5831, 2.0639, 2.3792,
  3.0284, 3.5237, 4.1868) of gauge invariant Gaussian smearing
~\cite{Gusken:1989qx} are applied symmetrically at the source (at $t=4$)
and at the sink. This is to ensure a variety of overlaps of the
interpolators with the lower-lying states. 
The analysis is performed on ten different quark masses corresponding to pion masses $m_{\pi}=\{0.797,0.729,0.641,0.541,0.430,0.380,0.327,0.295,\\0.249,0.224\}$ GeV.
 Error analysis is
performed using a second-order single elimination jackknife method,
where the ${\chi^{2}}/{\rm{dof}}$ is obtained via a covariance matrix
analysis method. Our fitting method is discussed extensively in Ref.~\cite{Mahbub:2009nr}.\\
The nucleon interpolator we consider is the local scalar-diquark
interpolator having a non-relativistic reduction ~\cite{Leinweber:1990dv,Leinweber:1994nm},
\begin{align}
\chi_1(x) &= \epsilon^{abc}(u^{Ta}(x)\, C{\gamma_5}\, d^b(x))\,
u^{c}(x).
\label{eqn:chi1_interpolator}
\end{align}

We consider several $4\times 4$ matrices. Each matrix is constructed with different sets of correlation functions, each set element corresponding to
a different numbers of sweeps of gauge invariant Gaussian smearing at the
source and sink of the $\chi_{1}\bar\chi_{1}$ correlators. This provides a
large basis of operators with varieties of overlap among energy
states. 
 \begin{figure*}[!hpt] 
  \begin{center}
   $\begin{array}{c@{\hspace{0.15cm}}c}  
 \includegraphics [height=0.48\textwidth,angle=90]{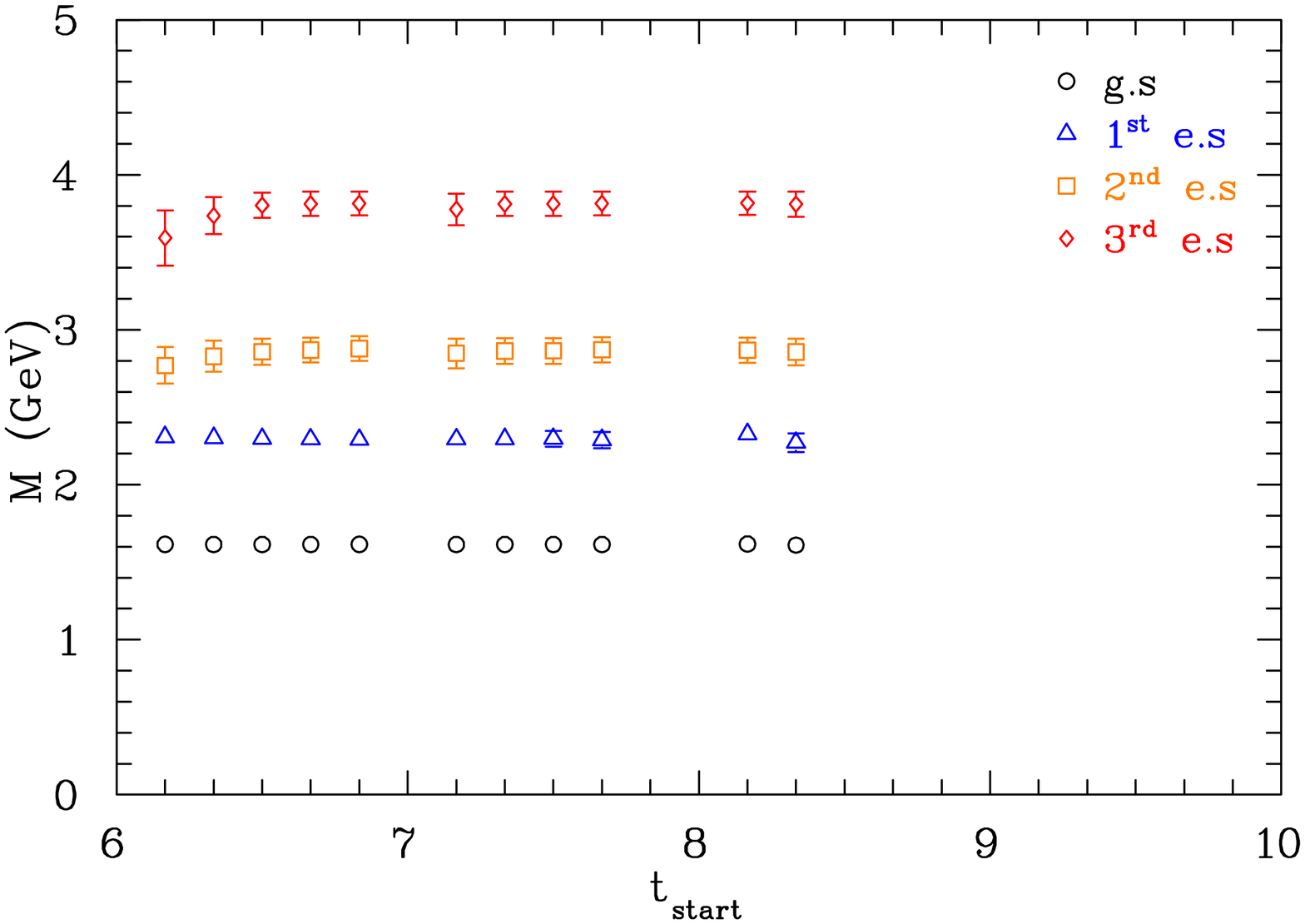} &
 \includegraphics [height=0.48\textwidth,angle=90]{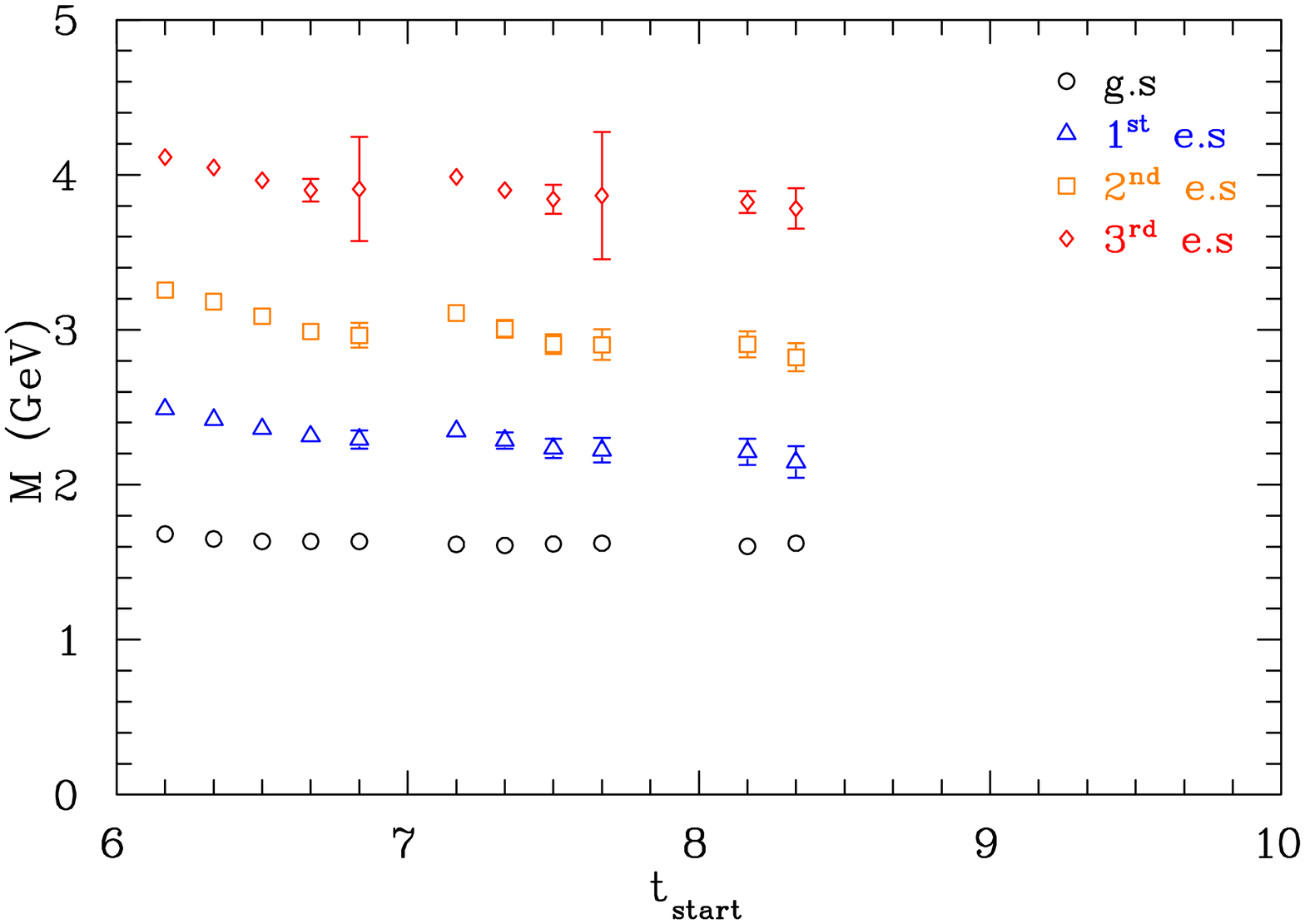}    
    \end{array}$
    \caption{(Color online). Masses of the nucleon, $N^{{\frac
          {1}{2}}^{+}}$ - states, from projected correlation functions as shown
      in Eq.(\ref{projected_cf_final}) (left figure) and from eigenvalues
      (right figure), for the pion mass of 797 MeV, and for the $4^{\rm{th}}$
      combination (3,12,26,35) of $4\times 4$ matrices. Each set of ground and
      excited states masses correspond to the diagonalization of the
      correlation matrix for each set of variational parameters
      $t\equiv t_{\rm start}$ (shown in major tick marks) and $\triangle t$
      (shown in minor tick marks). In the legend ``g.s'' stands for the ground
      state, whereas, ``e.s'' is for excited state. Larger values of
      $t_{\rm start}$ and $\triangle t$ did not provide a stable
      eigenvalue analysis.} 
   \label{fig:mass_and_eig_for_3.12.26.35_Q1}
  \end{center}
\end{figure*}
 \begin{figure*}[!hpt] 
  \begin{center}
 \includegraphics [height=0.80\textwidth,angle=90]{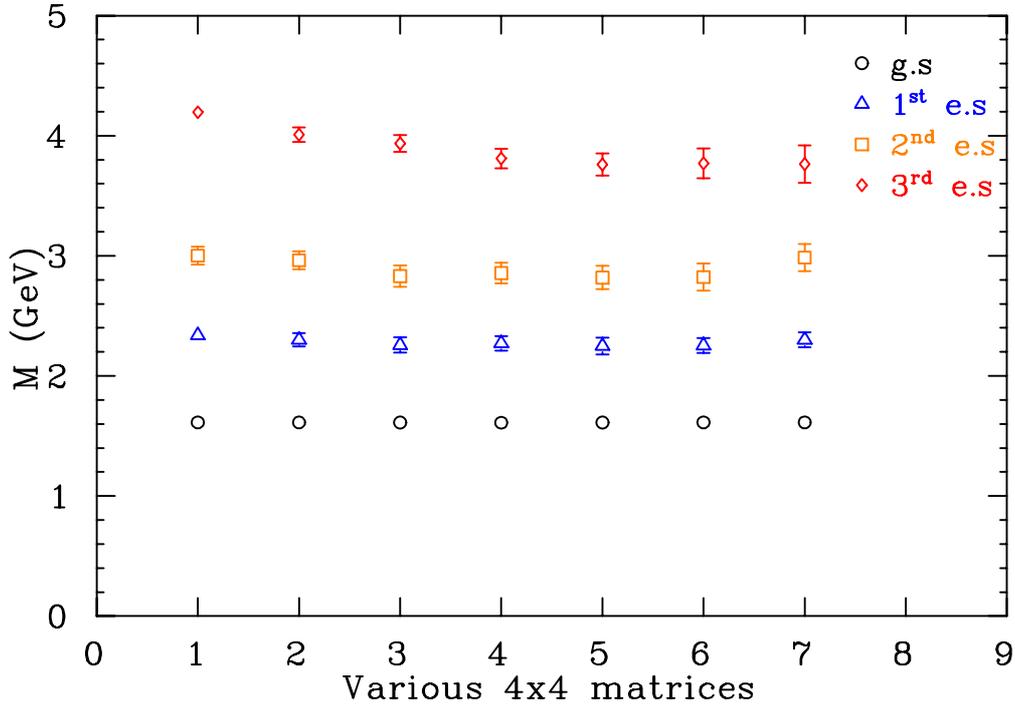}    
    \caption{(Color online). Masses of the nucleon, $N^{{\frac
          {1}{2}}^{+}}$ - states, from projected correlation functions as shown in Eq.(\ref{projected_cf_final})
      for the pion mass of 797 MeV. Numbers in
      the horizontal scale correspond to each combination of smeared $4\times 4$
      correlation matrices. For instance, 1 and 2 correspond to the combinations
      of (1,7,16,35) and (3,7,16,35) respectively and so on, as
      discussed in the text following Eq.(\ref{eqn:chi1_interpolator}). Masses are extracted according to the selection criteria described in
      Ref.~\cite{Mahbub:2009nr}.} 
   \label{fig:mass_for_all_combinations_Q1}
  \end{center}
\end{figure*}
We consider seven combinations \{1=(1,7,16,35), 2=(3,7,16,35),
3=(1,12,26,48), 4=(3,12,26,35), 5=(3,12,26,48),  6=(12,16,26,35),
7=(7,16,35,48)\} of $4\times 4$ matrices. 
In Ref.~\cite{Mahbub:2009nr} it was shown that one cannot isolate a
low-lying excited eigenstate using a single fixed-size source
smearing. The superposition of states manifested itself as a
smearing dependence of the effective mass. In this Letter we exploit
this sensitivity to isolate the energy eigenstates. 

In Fig.\ref{fig:mass_and_eig_for_3.12.26.35_Q1}, we show the mass from
the projected correlation functions and from the eigenvalues (as shown in
Ref.~\cite{Mahbub:2009nr}) for the fourth combination
(3,12,26,35) of $4\times 4$ matrices. We note that similar results in
mass from the projected correlation functions and the eigenvalues are
observed in this analysis as in Ref.~\cite{Mahbub:2009nr}. Though the
 mass of the first excited state from projected correlation functions
show little change with variational parameters, see
Fig.\ref{fig:mass_and_eig_for_3.12.26.35_Q1}, the second and third excited
states start a little below which indicates $t$ and $t+\triangle t$
are not sufficiently large. With larger Euclidean times fewer states
will contribute significantly to the correlators. The robust aspect of
fitting projected correlators is manifest in
Fig.\ref{fig:mass_and_eig_for_3.12.26.35_Q1}, and reflects the
stability of the eigenvectors against changes in $t$ and $t+\triangle t$.
In contrast, the mass from
the eigenvalue analysis shows significant dependence on the
variational parameters. 
The same method as described in
Ref.~\cite{Mahbub:2009nr} is applied in this Letter to
extract the mass from the projected correlation functions.

In Fig.\ref{fig:mass_for_all_combinations_Q1}, masses extracted from
all the combinations of $4\times 4$ matrices (from $1^{\rm st}$ to $7^{\rm
  th}$) are shown for the pion mass of 797 MeV. Some dependence of
the excited states on smearing count is also observed here as in
Ref.~\cite{Mahbub:2009nr} for a few of the interpolator basis smearing
sets. However the ground and first excited states are robust against
changes in the interpolator basis, providing evidence that an energy
eigenstate has been isolated. It should be noted
that the highest excited state (the third excited state) is influenced more
by the level of smearing than the lower excited states. This is to be
expected as this state must accommodate all remaining spectral strength.    

The $1^{\rm{st}}$ combination in
Fig.\ref{fig:mass_for_all_combinations_Q1} provides heavier excited
states as this basis begins with a low number of smearing sweeps (a
sweep count of 1) and also contains another low smearing set of 7
sweeps. The second and third excited states, and more importantly, the
first excited state sits a bit high in comparison with the other
bases. Hence, extracting masses with this basis is not as reliable as
other sets. The
$2^{\rm{nd}}$ combination also contains elements with a small smearing
sweep count (3 and 7), hence this basis also provides heavier excited
states and shows some systematic drift in the second excited
state. However, this basis has reduced contamination from the excited
states when compared with the first basis. The $3^{\rm rd}$
combination also starts at the low smearing count, so the mass from
this basis for the third excited state is a little high.

We can observe at this point that including basis elements with a low
smearing count will increase the masses of excited states (for
instance, consecutive low numbers of smearing sweeps 1,7 and 3,7,
respectively). This is because the correlation functions with these
low sweep counts have a large overlap with several heavier excited states in
their sub-leading exponential. We also observe that
the inclusion of basis elements with a high level of smearing (for
instance, a sweep count of 48) results in larger statistical errors in
the analysis.

The $6^{\rm{th}}$ combination starts from a moderate smearing sweep
count (12) and also contains elements with consecutive smearing sweep
counts which provides less diversity in the basis. Similarly, the
$7^{\rm th}$ combination
contains consecutive large smearing sweep counts of 35 and 48 and
operators for these levels of smearing are very similar challenging
the isolation of single energy eigenstates. 

The $4^{\rm th}$ and $5^{\rm th}$ combinations are well spread over
the given range of smearing sweeps and start from a sweep count of 3.
They don't include successive lower smearing sweep counts. The $5^{\rm
  th}$ combination contains the basis element with a sweep count of 48
but has only slightly larger statistical errors than the $4^{\rm th}$
basis choice. Both these bases provide diversity.
 It is observed that both combinations
provide consistent results for the states.
 Nonetheless, it should be noted that the $3^{\rm rd}$ to
$6^{\rm th}$ combinations all provide very consistent results for the
lower three (ground, first and second) energy states, shown in
Fig.\ref{fig:mass_for_all_combinations_Q1}. While the $2^{\rm{nd}}$
and the $3^{\rm{rd}}$ excited states display some evidence of
eigenstate mixing, the ground and the $1^{\rm st}$
excited states are robust in this analysis as they agree within one
standard deviation for all the combinations of $4\times 4$ matrices.
Hence, an
analysis is performed to calculate the systematic errors associated
with the choice of basis over the preferred four combinations (from
$3^{\rm rd}$ to $6^{\rm th}$) with
$\sigma_{b}=\sqrt{{\frac{1}{N_{b}-1}}\sum_{i=1}^{N_{b}}(M_{i}-\bar{M})^{2}}$,
where, $N_{b}$ is the number of bases, in this case is equal to 4. It
should be noted that for the ground and first excited states the systematic
errors associated with choice of basis are very small in
comparison with their statistical errors (see Table \ref{table:table.mpi.mavg.err-sum-over-qudrature.4comb.4states}).

 \begin{figure*}[!hpt] 
  \begin{center} 
 \includegraphics [height=0.96\textwidth,angle=90]{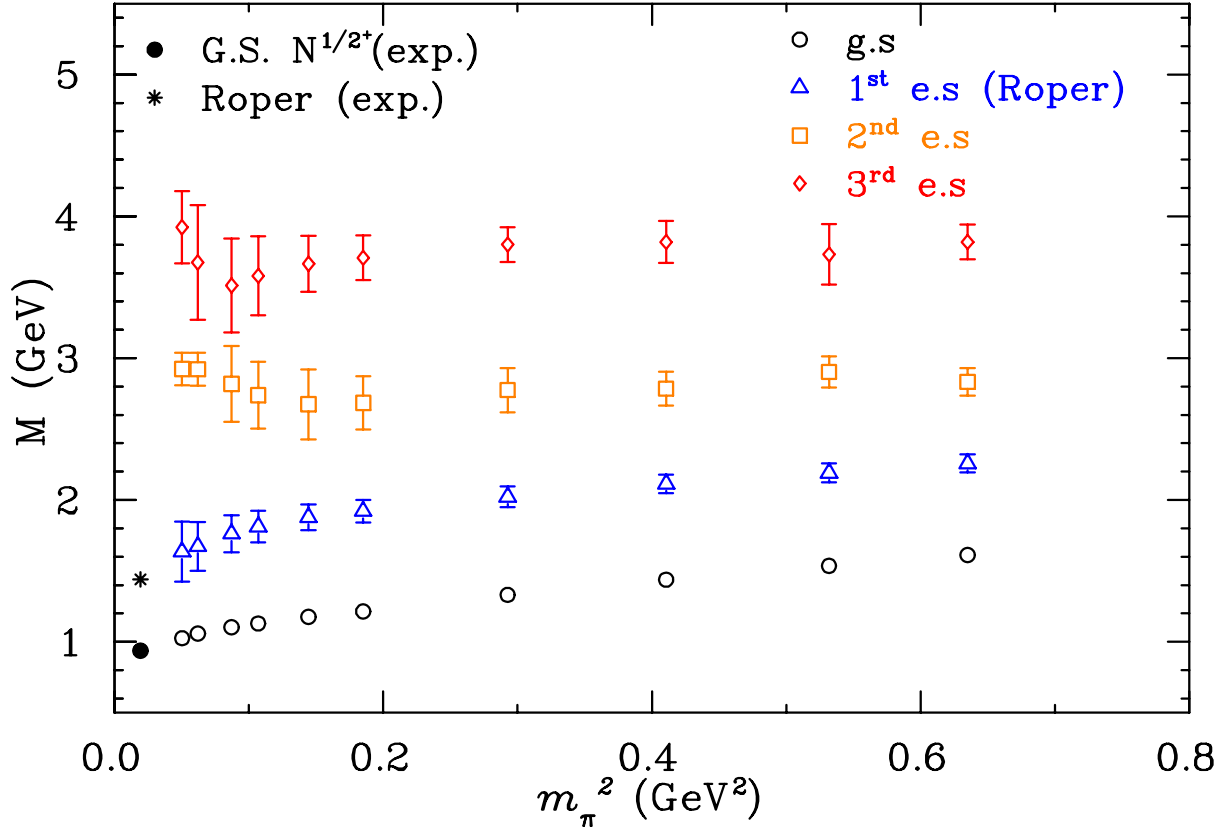}    
    \caption{(Color online). Mass of the nucleon,
      $N^{{\frac{1}{2}}^{+}}$ - states, for the ground and the excited states. The
      errors shown in the figure are a combination of average statistical errors
      and systematic errors due to basis
      choices over four bases (from $3^{\rm rd}$ to
      $6^{\rm th}$ of Fig.\ref{fig:mass_for_all_combinations_Q1}). Errors
      are combined in quadrature. The
      black filled symbols are the experimental values of the ground
      and the Roper states of the nucleon ~\cite{Yao:2006px}. The rightmost point corresponds to the pion mass of
      797 MeV, then for 739, 641, 541, 430, 380, 327, 295, 249 and 224
      MeV (leftmost point).} 
   \label{fig:paper_m.avg_4com.x1x1.4x4.sqrt_erravg_4combs_errbasis_4combs.4states.allQ.NEW}
  \end{center}
\end{figure*}
  \begin{table*}[!hpt]
    \begin{center}
    \caption{\label{table:table.mpi.mavg.err-sum-over-qudrature.4comb.4states}
      Mass of the nucleon, $N^{{\frac{1}{2}}^{+}}$ - states, are averaged over
      the four bases (from $3^{\rm rd}$ to $6^{\rm th}$) and the
      errors for the nucleon shown here are a combination of average statistical  errors and systematic errors for choice of
      basis over the four bases. The $2^{\rm nd}$ and $3^{\rm rd}$ states may be superposition of energy eigenstates as discussed in the text.}
   \vspace{0.5cm}
    \begin{tabular}{p{2.5cm}p{2.5cm}p{2.5cm}p{2.5cm} p{1.25cm}} 
        \hline
        \hline
   $aM_{\pi}$ & $aM^{N^{{\frac{1}{2}}^{+}}}_{g.s}$ & $aM^{N^{{\frac{1}{2}}^{+}}}_{1^{\rm st}\  \rm{e.s}}$(Roper) & $aM^{N^{{\frac{1}{2}}^{+}}}_{2^{\rm nd}\ \rm{e.s}}$ & $aM^{N^{{\frac{1}{2}}^{+}}}_{3^{\rm rd}\ \rm{e.s}}$ \\
        \hline 
 0.5141(19) & 1.0399(65) & 1.457(41) & 1.827(62)  & 2.464(78) \\   
 0.4705(20) & 0.9905(71) & 1.413(43) & 1.872(71)  & 2.40(13)  \\
 0.4134(22) & 0.9280(79) & 1.364(42) & 1.797(77)  & 2.464(95) \\
 0.3490(24) & 0.8589(90) & 1.304(46) & 1.79(10)   & 2.452(78) \\
 0.2776(24) & 0.783(11)  & 1.239(51) & 1.73(12)   & 2.39(10)  \\
 0.2452(24) & 0.758(13)  & 1.212(58) & 1.72(15)   & 2.36(12)  \\
 0.2110(27) & 0.728(13)  & 1.169(71) & 1.76(15)   & 2.30(17)  \\
 0.1905(31) & 0.711(12)  & 1.137(83) & 1.81(17)   & 2.26(21)  \\
 0.1607(35) & 0.682(13)  & 1.07(11)  & 1.885(74)  & 2.37(26)  \\
 0.1448(44) & 0.661(15)  & 1.05(13)  & 1.885(73)  & 2.53(16)  \\

\hline
\hline
  \end{tabular}
 \end{center}
 \end{table*}

  \begin{table*}[!hpt]
    \begin{center}
    \caption{\label{table:mass_RoperStates_4combs} Mass of the Roper state,
      $N^{{\frac{1}{2}}^{+}}$, for four bases \{(1,12,26,48), (3,12,26,35),
      (3,12,26,48) and (12,16,26,35)\} of $4\times 4$ matrices.
      $t_{1}$
      and $t_{2}$ represent the lower and upper times respectively of
      fit windows in the projected effective mass. Inclusion of $t_{1}=6$ produces $\chi^{2}/{\rm dof}\gg 1$.}

    \begin{tabular}{p{0.15cm}p{.4cm}p{1.4cm}p{.6cm} | p{.15cm}p{.4cm}p{1.4cm}p{.6cm} | p{0.15cm}p{.4cm}p{1.4cm}p{.6cm} | p{0.15cm}p{.4cm}p{1.4cm}p{.6cm}} 
        \hline
        \hline
    \multicolumn{4}{c|}{$3^{\rm rd}$ basis (1,12,26,48)} & \multicolumn{4}{c|}{$4^{\rm th}$ basis (3,12,26,35)} & \multicolumn{4}{c|}{$5^{\rm th}$ basis (3,12,26,48)} & \multicolumn{4}{c}{$6^{\rm th}$ basis (12,16,26,35)} \\
        \hline 
 $t_{1}$ & $t_{2}$ & $aM$ (Roper) & $\frac{\chi^{2}}{\rm dof}$ & $t_{1}$ & $t_{2}$ & $aM$ (Roper) & $\frac{\chi^{2}}{\rm dof}$ & $t_{1}$ & $t_{2}$ & $aM$ (Roper) & $\frac{\chi^{2}}{\rm dof}$ & $t_{1}$ & $t_{2}$ & $aM$ (Roper) & $\frac{\chi^{2}}{\rm dof}$ \\ 
\hline
 7 & 12 & 1.456(41) & 0.58 & 7 & 12 & 1.465(39) &  0.63 &  7 & 12 &  1.451(44) & 0.51 & 7 & 12 & 1.454(40) & 0.57  \\
 7 & 12 & 1.411(43) & 0.55 & 7 & 12 & 1.419(41) &  0.62 &  7 & 12 &  1.405(46) & 0.48 & 7 & 12 & 1.417(39) & 0.60  \\
 7 & 12 & 1.368(39) & 0.54 & 7 & 12 & 1.361(45) &  0.60 &  7 & 12 &  1.364(40) & 0.53 & 7 & 11 & 1.363(42) & 0.68  \\
 7 & 12 & 1.307(44) & 0.57 & 7 & 11 & 1.298(51) &  0.60 &  7 & 12 &  1.305(45) & 0.57 & 7 & 10 & 1.308(46) & 0.54  \\
 7 & 11 & 1.235(50) & 0.43 & 7 & 11 & 1.245(51) &  0.57 &  7 & 11 &  1.233(51) & 0.37 & 7 & 11 & 1.244(52) & 0.38  \\
 7 & 11 & 1.210(60) & 0.42 & 7 & 11 & 1.211(55) &  0.58 &  7 & 11 &  1.206(57) & 0.38 & 7 & 11 & 1.220(60) & 0.49  \\
 7 & 10 & 1.163(69) & 0.60 & 7 & 11 & 1.165(67) &  0.56 &  7 & 10 &  1.164(71) & 0.53 & 7 & 10 & 1.184(75) & 0.56  \\
 7 & 10 & 1.129(82) & 0.61 & 7 & 10 & 1.127(81) &  0.84 &  7 & 10 &  1.136(82) & 0.58 & 7 & 10 & 1.155(85) & 0.54  \\
 7 & 10 & 1.07(10)  & 0.56 & 7 & 10 & 1.06(10)  &  0.95 &  7 & 10 &  1.07(11)  & 0.68 & 7 & 10 & 1.11(11)  & 0.63  \\
 7 &  9 & 1.04(13)  & 0.85 & 7 & 10 & 1.01(12)  &  0.97 &  7 & 9  &  1.05(13)  & 0.79 & 7 &  9 & 1.10(13)  & 0.70  \\
\hline
\hline
  \end{tabular}
 \end{center}
 \end{table*}

\begin{figure*}[!hpt] 
  \begin{center} 
 \includegraphics [height=0.96\textwidth,angle=90]{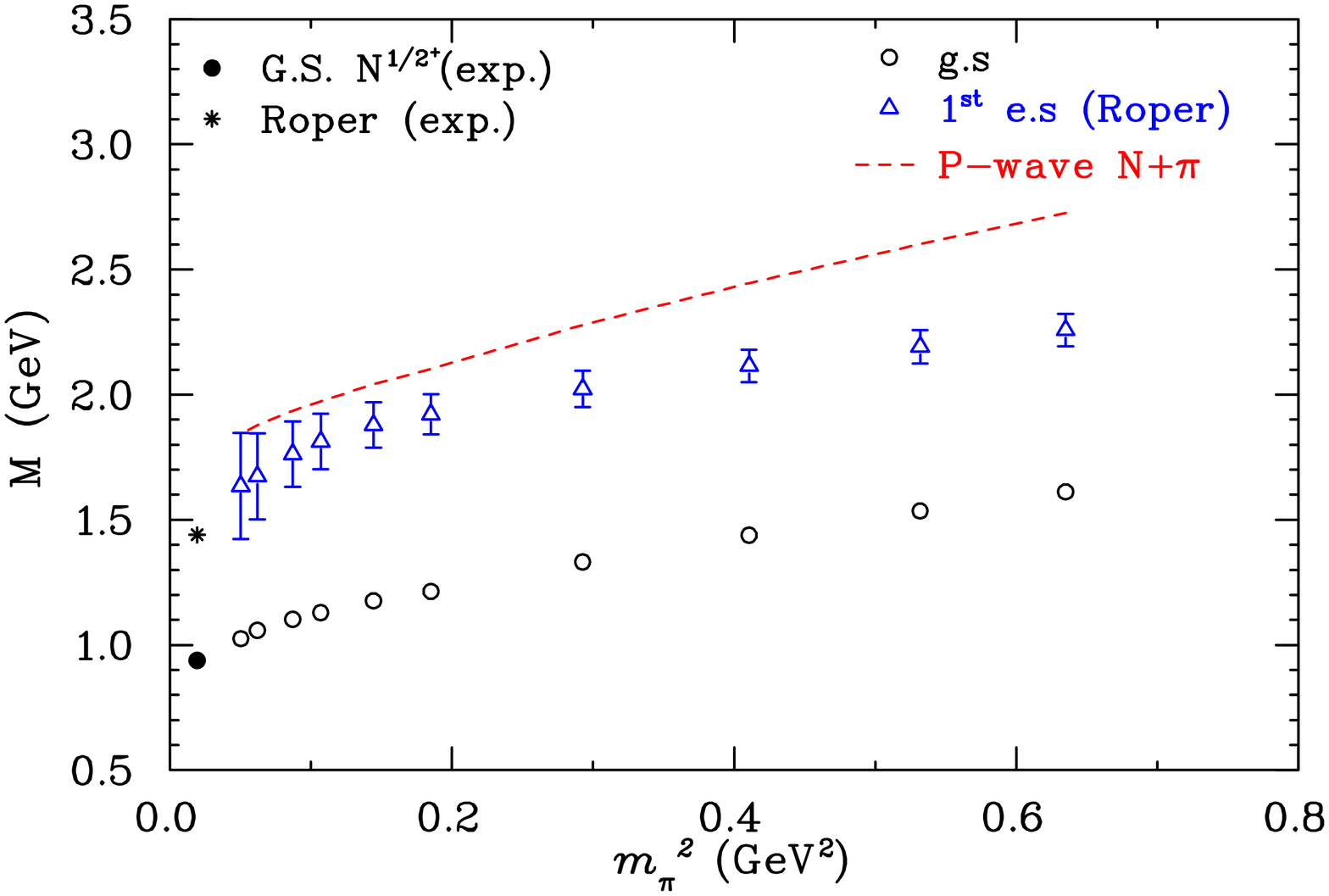}    
    \caption{(Color online). The ground and the Roper states
      and the non-interacting
      P-wave $N+\pi$ are illustrated. The black filled symbols are the
      experimental values of the ground and the Roper states obtained
      from Ref.~\cite{Yao:2006px}.} 
   \label{fig:paper_m.avg_4com.x1x1.4x4.sqrt_erravg_4combs_errbasis_4combs.2states.allQ.NEW}
  \end{center}
\end{figure*}
In
Fig.\ref{fig:paper_m.avg_4com.x1x1.4x4.sqrt_erravg_4combs_errbasis_4combs.4states.allQ.NEW},
the masses from projected correlation functions are shown for all
pion masses averaged over four combinations of correlation matrices (from
$3^{\rm rd}$ to $6^{\rm th}$). Masses are averaged over these
four bases and errors (average statistical errors over these four
bases and systematic errors associated with basis choices) are
combined in quadrature,
$\sigma=\sqrt{\bar{\sigma}_{s}^{2}+\sigma_{b}^{2}}$.
Eigenstate mixing may be affecting the results for
the second and the third excited states. The ground and the
$1^{\rm{st}}$ excited states systematically approach the physical mass
of the Nucleon and Roper state. This is the first time
evidence of a low-lying Roper state has appeared from variational
analysis near the chiral regime.

In
Fig.\ref{fig:paper_m.avg_4com.x1x1.4x4.sqrt_erravg_4combs_errbasis_4combs.2states.allQ.NEW},
the ground and the first excited states of
Fig.\ref{fig:paper_m.avg_4com.x1x1.4x4.sqrt_erravg_4combs_errbasis_4combs.4states.allQ.NEW}
are shown in larger scale for clarity. The non-interacting two
particle P-wave
$N+\pi$ is shown by the dashed line. It is interesting
to note that the observed lattice Roper state sits lower than the P-wave
$N+\pi$, indicative of attractive $\pi {\rm N}$ interactions producing
a resonance at physical quark masses. The very consistent Euclidean-time fit window for all the
bases and for all the quark masses, noticeably the lower time of the
fit as shown in Table \ref{table:mass_RoperStates_4combs}, assist to identify the robust nature of the extracted Roper state which
comes at the same Euclidean-time from the heaviest to the lightest
quarks masses.

There are quenched artifacts that will have significant influence on our
results should we progress to lighter quark masses. Not only are $\pi N$ couplings different in quenched QCD, but there are also
contributions from unphysical, degenerate P-wave $\eta^{\prime} N$
  two-particle states ~\cite{Mathur:2003zf}. Future calculations
  approaching the decay threshold should be done in full dynamical-fermion QCD.

In conclusion, through the use of a variety of smeared-smeared
correlation functions in constructing the 
correlation matrix, the first positive parity excited state of the nucleon
$N^{{\frac{1}{2}}^{+}}$, the Roper state, has been observed for the
first time using the variational analysis. While the $3\times 3$
correlation  matrix
analysis of standard interpolators is insufficient to isolate these
excited energy eigenstates of QCD ~\cite{Mahbub:2009nr}, using the new correlation matrix construction with
smeared-smeared correlators enables us to extract the otherwise
elusive Roper state.

\section*{Acknowledgments}
We thank the NCI National Facility and eResearch SA for generous
grants of supercomputing time which have enabled this project.  This
research is supported by the Australian Research Council.
\end{document}